\documentclass[10pt, twocolumn, oneside]{article}
\usepackage{graphicx} 
\usepackage{color} 
\usepackage[normalem]{ulem} 
\usepackage{multirow} 
\usepackage{amsfonts} 
\usepackage{amsmath,amssymb}
\usepackage{hyperref}
\usepackage{extsizes}
\usepackage{physics} 
\usepackage{comment} 
\usepackage{bm} 
\usepackage{xcolor} 
\usepackage{authblk} 
\usepackage[sort&compress,numbers]{natbib}
\usepackage{dcolumn}
\usepackage{bm}
\usepackage{mathtools}

\newcommand{\iu}{\mathrm{i}\mkern1mu}

\newcommand{\lhat}{\mathbf{e}_L}
\newcommand{\rhat}{\mathbf{e}_R}
\newcommand{\zhat}{\mathbf{e}_z}
\newcommand{\xhat}{\mathbf{\hat{x}}}
\newcommand{\yhat}{\mathbf{\hat{y}}}
\newcommand{\ii}{\mathrm{i}}

\usepackage{tocloft}
\hypersetup{
  colorlinks,
  citecolor=blue!50!green!90!black,
  linkcolor=blue!80!green!90!black,
  urlcolor=blue!80!green!90!black}

\usepackage[
    left=15mm, 
    right=15mm, 
    top=23mm,
    bottom=23mm, 
    columnsep=9mm 
]{geometry}

\title{
Light structuring via nonlinear total angular momentum addition \\ with flat optics}

\author[1]{Evgenii Menshikov\thanks{Email: evgenii.menshikov@unibs.it}}
\author[1,2]{Paolo Franceschini}
\author[1]{Kristina Frizyuk\thanks{Email: frzyuk@gmail.com}}
\author[3]{Ivan Fernandez-Corbaton}
\author[4,2]{Andrea Tognazzi}
\author[4]{Alfonso Carmelo Cino}
\author[7]{Denis Garoli}
\author[5]{Mihail Petrov}
\author[1, 2]{Domenico de Ceglia}
\author[1, 2]{Costantino De Angelis\thanks{Email: costantino.deangelis@unibs.it}}

\affil[1]{Department of Information Engineering, University of Brescia, Via Branze, 38, Brescia 25123, Italy}
\affil[2]{National Institute of Optics-National Research Council, Via Branze, 45, Brescia, 25123, Italy}
\affil[3]{Karlsruhe Institute of Technology, Kaiserstrasse, 12, Karlsruhe, 76131, Germany}
\affil[4]{Department of Engineering, University of Palermo, Viale delle Scienze, 9, Palermo, 90128, Italy}
\affil[5]{Qingdao Innovation and Development Center, Harbin Engineering University, Sansha road, 1777, Qingdao, 266000, China}
\affil[7]{{Dipartimento di Scienze e Metodi dell’Ingegneria, Università degli Studi di Modena e Reggio Emilia, Via Amendola, 2, Reggio Emilia, 43122, Italy }}
\date{} 

\begin{document}

\maketitle
\begin{abstract}
   {Shaping the structure of light with flat optical devices has driven significant advancements in our fundamental understanding of light and light-matter interactions, and enabled a broad range of applications, from image processing and microscopy to optical communication, quantum information processing, and the manipulation of microparticles. 
   Yet, pushing the boundaries of structured light beyond the linear optical regime remains an open challenge. Nonlinear optical interactions, such as wave mixing in nonlinear flat optics, offer a powerful platform to unlock new degrees of freedom and functionalities for generating and detecting structured light.  
   In this study, we experimentally demonstrate the non-trivial structuring of third-harmonic light enabled by the addition of total angular momentum projection in a nonlinear, isotropic flat optics element --- a single thin film of amorphous silicon. 
   We identify the total angular momentum projection and helicity as the most critical properties for analyzing the experimental results. The theoretical model we propose, supported by numerical simulations, offers quantitative predictions for light structuring through nonlinear wave mixing under various pumping conditions, including vectorial and non-paraxial pump light. 
   Notably, we reveal that the shape of third-harmonic light is highly sensitive to the polarization state of the pump. Our findings demonstrate that harnessing the addition of total angular momentum projection in nonlinear wave mixing can be a powerful strategy for generating and detecting precisely controlled structured light.}
\end{abstract}

\vspace{.5cm}
\tableofcontents

\section{Introduction}
Light beams with carefully engineered spatial structures can carry various transverse modes, offering an extraordinary resource when harnessed effectively~\cite{Forbes2021-Structuredlight}.
This has ignited a surge of interest in the field of structured light, leading to numerous applications across diverse areas such as image processing~\cite{Davis:00, furhapter2005spiral},
superresolution microscopy \cite{klar2001breaking, kozawa2018superresolution, Yoshida:19},
metrology \cite{d2013photonic},
communication~\cite{wang2012terabit},
quantum information processing~\cite{vallone2014free, nicolas2014quantum},
and light–matter interactions~\cite{garcia2024topological, padgett2011tweezers}. 
Furthermore, structured light in nonlinear optical devices shows great promise for advancing analog deep neural networks, where the vast number of modes can be leveraged to enhance parallel processing and data encoding capabilities~\cite{Lin2018-All-opticalmachinel, Wang2024-Large-scalephotonic}. 
Recently, a relevant paradigm shift has occurred in the light shaping field, driven by advancements in flat optics and photonic nanostructures~\cite{devlin2017, dorrah2022}.

To fully exploit the potential in the above-mentioned applications, a powerful and yet simple approach to describe the input--output relationships in real devices is needed. 
Very often, to simplify this task, which involves vector fields, one tries to construct some scalar description of the problem. 
In some cases, however, considering the vector field as a whole can actually simplify matters when one exploits the symmetry properties of both the light field and the material systems. 
For example, considering the behavior of the fields under rotations about a particular axis is relevant for samples with cylindrical symmetry. 
Such behavior is determined by the projection of the total angular momentum (TAM) of the field on the optical axis~\cite{Akhiezer}.  
Another property of the fields that helps in the analysis of experiments~\cite{FerCor2012b,tischler2014experimental,Zambrana2016} is the handedness of light, or helicity~\cite{Zwanziger1968, Weinberg2005-TheQuantumTheoryof, Tung2020-GroupTheoryInPhysi}. 
Both total angular momentum and helicity are fundamental quantities of the electromagnetic field, which are connected to the symmetries of Maxwell's equations (see Suppl. Inf.~\ref{sec:definitions} for definitions). 
Their analysis allows to readily predict some aspects of the light-matter interaction using the conservation laws related to the particular symmetries of the system. 
Both quantities are generally valid for electromagnetic fields, independently of whether the fields are collimated, focused, far-fields, or evanescent.



There are well-known examples of fields that transform under rotations in particularly simple ways: the vector spherical harmonics, which are spherical waves describing the electric and magnetic fields of multipoles~\cite{Stratton, Stein1961-Additiontheoremsfor, Akhiezer} (well-defined TAM and TAM projection), or the Bessel beams \cite{Hacyan2006}, which are cylindrical waves (well-defined TAM projection). 
With respect to helicity, one may obtain purely circularly polarized (CP) spherical or cylindrical waves by linear combinations of electric and magnetic multipolar fields~\cite{Tung2020-GroupTheoryInPhysi}, or TE/TM Bessel beams~\cite{Afanasev2013}.

\begin{figure}[t]
    \centering
    \includegraphics[width=\columnwidth]{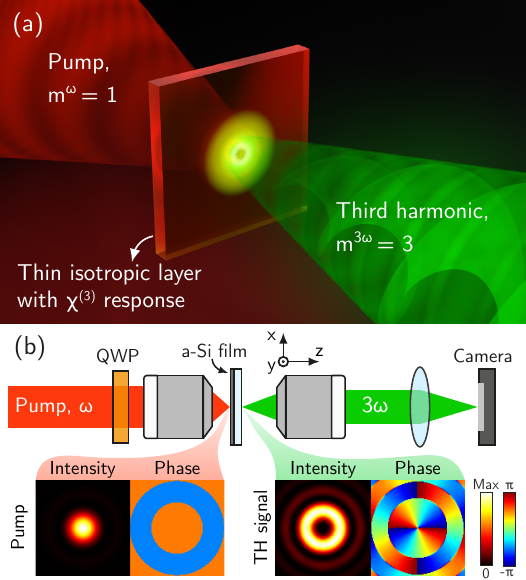}
    \caption{(a) Illustration of the nonlinear total angular momentum (TAM) $z$-projection ($m$) addition process. 
    When a circularly polarized (CP) pump beam ($m^{\omega} = 1$) is incident on a thin isotropic layer with $\chi^{(3)}$ nonlinearity, the generated signal at third-harmonic (TH) frequency has tripled TAM ($m^{3\omega} = 3$). 
    (b) Schematic of the optical setup for observation of the non-linear TAM projection addition in a thin film of a-Si. 
    Bottom panel shows numerical simulations of the intensity, $|\vb{E}\cdot \vb{e}_R^*|^2$, and phase, $\text{arg}(\vb{E}\cdot \vb{e}_R^*)$, distributions of the dominant polarization components in the right CP pump and in the generated TH signal on the thin film surface.} 
    \label{thg_illust}
\end{figure}

In this work, we experimentally demonstrate the tripling of the total angular momentum projection through third-order nonlinearity in an amorphous silicon flat film as schematically shown in Fig.~\ref{thg_illust}. 
All experimental findings are supported by a theoretical framework and numerical simulations.

The rest of the article is organized as follows.  
In Section~\ref{sec:experiment}, we present the results of experimental measurements of third-harmonic signal distributions obtained under tightly focused excitation with various polarization parameters. 
Section~\ref{sec:sec3} presents the theoretical framework that describes our findings. 
First, we introduce a straightforward and comprehensive explanation leveraging the rotational symmetry of our system. 
After that, we provide a numerical model that produces results closely aligned with the experimental observations.

\section{Experimental results}
\label{sec:experiment}
We study third-harmonic (TH) generation in amorphous silicon (a-Si) thin films deposited on a fused silica (SiO$_2$) substrate (see Methods~\ref{sec:methods} for details). 
The choice of the substrate material is motivated by its transparency at both the pump and TH wavelengths, and relatively low nonlinear properties, allowing one to neglect the contribution of the substrate to the detected TH signal ($\chi^{(3)}$ of a-Si is 4 orders of magnitude larger than that of SiO$_2$ \cite{ikeda2007enhanced, Boyd2003}). 
\begin{figure*}[t]
    \centering
    \includegraphics[width=\textwidth]{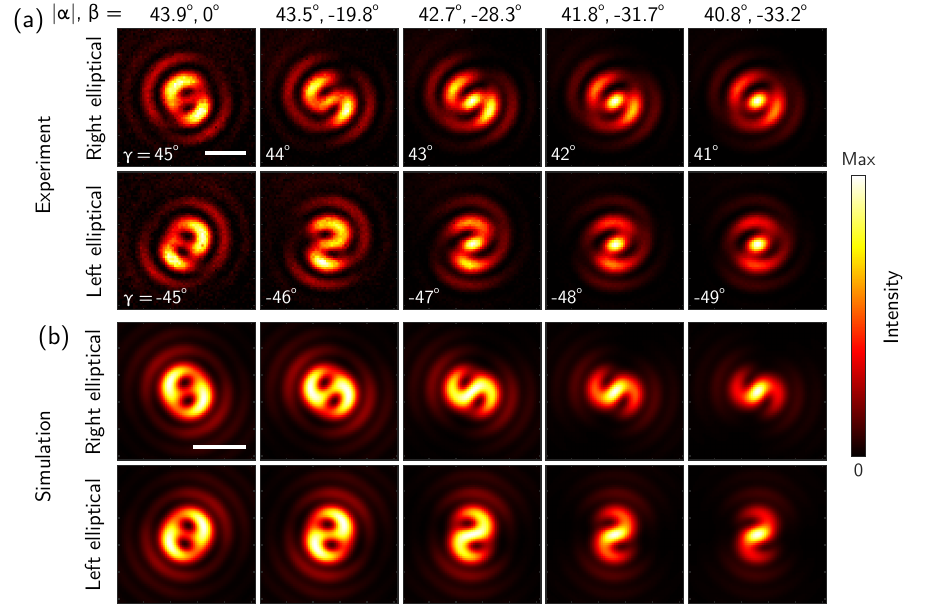}
    \caption{(a) Optical images of patterns at TH excited with an elliptically polarized pump laser. Here, $\alpha$ and $\beta$ denote the ellipticity angle and the inclination angle of the polarization ellipse, respectively, for the pump beam entering the objective lens; $\gamma$ is the angle between the fast axis of the QWP and the polarization direction of input laser beam. 
    The deviation of the retardance of the used QWP from perfect limits the maximum ellipticity angle of the pump to $43.9^\circ$. 
    (b) Numerical simulation of patterns generated by a thin isotropic layer with $\chi^{(3)}$ nonlinearity when illuminated with a tightly focused laser beam. Scale bars, 2 $\mu$m.} 
    \label{spiral_exp&sim}
\end{figure*}
A simplified schematic of our optical setup is shown in Fig.~\ref{thg_illust}b. Here, a pump beam with a wavelength of 1500 nm is focused on the surface of the a-Si film using an objective lens with high numerical aperture ($\text{NA}=0.85$). 
The polarization of the pump beam can be adjusted by changing the orientation of the fast axis of a quarter-wave plate (QWP) mounted on a motorized rotation stage, allowing for excitation with polarizations ranging from linear to nearly circular. 
The generated TH signal is collected by an objective lens and imaged onto the camera sensor using a tube lens.

Figure~\ref{spiral_exp&sim}a shows optical images of patterns at TH, obtained for decreasing angles $\gamma$ between the fast axis of the wave plate and the polarization direction of the input laser beam. This corresponds to decreasing absolute values of the ellipticity angle $\alpha$ of the pump beam entering the objective lens and the angle of the polarization ellipse inclination $\beta$. The angles are defined through the Stokes parameters as $2\alpha = \asin(S_3/S_0)$, and $2\beta = \atan(S_2/S_1)$~\cite{collett1992polarized}. 
We find out that irradiating a thin a-Si film by a tightly focused pump beam with polarization close to circular, results in generation of a TH signal with two minima in its intensity distribution (see Fig.~\ref{spiral_exp&sim}a, ellipticity angle $43.9^\circ$). Here we start with a maximum achievable value of $\alpha_0 = 43.9^\circ$ (vs $45^\circ$ for a pure CP light) due to deviation of the QWP retardance from a perfect one.
As the ellipticity angle decreases, we observe the formation of a spiral pattern with two lobes, which then transforms into a shape with one maximum at the center of the pattern. 

The experimental results can be well reproduced in the numerical simulations shown in Fig.~\ref{spiral_exp&sim}b. 
Before describing the details of the numerical simulations in Section~\ref{sec:simulation}, we show that the experimental results can be understood through the conservation of the TAM projection on the optical axis. 

\begin{table*}
    \centering
    \begin{tabular}{|c|c|c|c|}
        \hline
		 & $\rhat=(\xhat+\ii\yhat)/\sqrt{2}$ & $\lhat=(\xhat-\ii\yhat)/\sqrt{2}$ & $\zhat$\\
        \hline
		$(m,\lambda)$ & $J_{m-1}\exp(\ii(m-1)\varphi)$ & $J_{m+1}\exp(\ii(m+1)\varphi)$ & $J_{m}\exp(\ii\varphi m)$\\
		$\theta_{\text{max}}\rightarrow 0$, $(m,{1})$  & $\rightarrow J_{m-1}\exp(\ii(m-1)\varphi)$ & $\rightarrow 0$ & $\rightarrow 0$\\
		 $\theta_{\text{max}}\rightarrow 0$, $(m,{-1})$ & $\rightarrow 0$  & $\rightarrow J_{m+1}\exp(\ii(m+1)\varphi)$& $\rightarrow 0$\\
        \hline
    \end{tabular}
    \caption{With $\varphi=\atan(y,x)$, the first row shows the phase singularities attached to each polarization of a $(m,\lambda)$ beam of well defined $\text{TAM}=m$ and helicity $\lambda=\pm {1}$. 
    The argument of the Bessel functions $J_n(\cdot)$ is suppressed, as it changes with $\theta$ in Eq.~\eqref{eq:bbexp}. 
    The second and third rows show the dominant polarization component in the collimated limit (see Suppl. Inf.~\ref{app:BBs}). 
    } 
    \label{tab:BB}
\end{table*}
\begin{figure*}[h!]
    \centering
    \includegraphics[width=\textwidth]{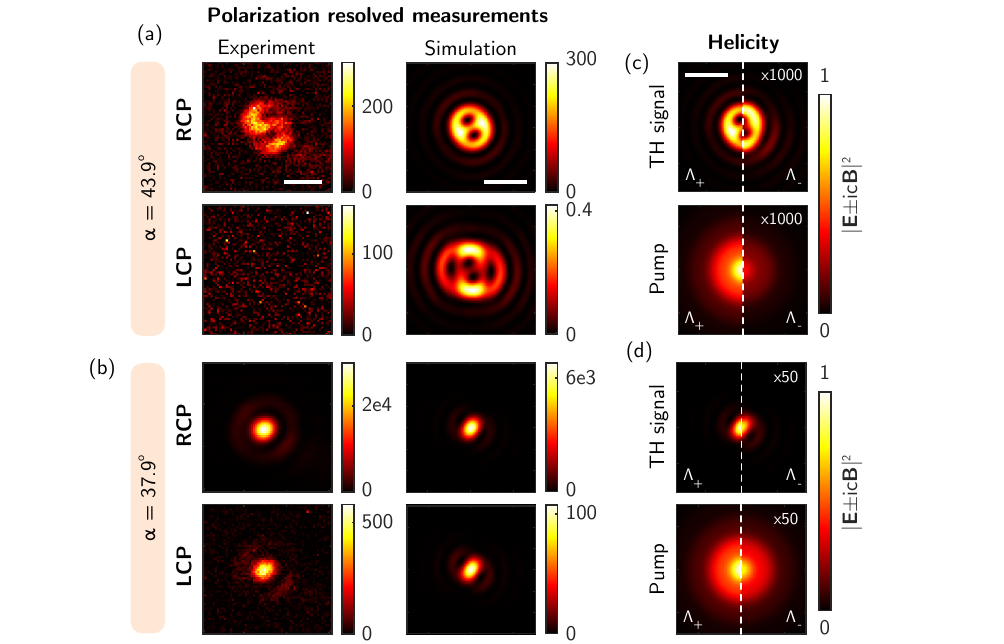}
    \caption{
    Optical images and numerical simulations of polarization resolved patterns at TH excited with an elliptically polarized pump laser with $\alpha = 43.9^{\circ}$ (a)  and $37.9^{\circ}$ 
    (b) Numerical simulations of helicity of the pump and generated TH signal for $\alpha = 43.9^{\circ}$ 
    (c) and $37.9^{\circ}$ 
    (d) Electric fields were calculated on the film surface, helicity values are normalized by the corresponding maximum value for $\Lambda_+$. Scale bars, 2 $\mu$m.} 
    \label{polar_res}
\end{figure*}
\section{Discussion}
\label{sec:sec3}
\subsection{Theoretical setting}
\label{sec:theorySi}
As shown in Fig.~\ref{thg_illust}, the material system is invariant under rotations along the optical axis when considered from right after the QWP up to the camera. 
Crucial for such invariance is the isotropy of the a-Si where the TH is generated. 
Symmetry then dictates that the TAM $z$-projection ($m$) on the optical axis is conserved. 
The precise meaning of this conservation in non-linear interactions deserves clarification. 
Let us assume that the incident pump light has TAM projection $m^{\omega}=1$. Then, since three photons of the incident beam are combined in the material, the TH output must have TAM projection $m^{3\omega}=3$. 
So, the conservation due to symmetry leads to tripling in this case. 
Such kinds of predictions can be rigorously obtained using the expressions of the angular momentum operators in tensor product spaces, as done in~\cite[Sec.~III-C]{Freter2024} for second harmonic generation. 
However, when the input contains a superposition of more than one component of $m^{\omega}$, e.g., 1 and $-1$, the same conservation law implies that each possible combination of three instances of the input components will produce a TH response with $m^{3\omega}$ equal to the sum of the three contributions. 
For example, in the top row of Fig.~\ref{spiral_exp&sim}, the incident beam has evolved from an almost pure content of $m^{\omega}=1$ to a mix, in which the intensity of an additional $m^{\omega}=-1$ component grows as the ellipticity angle of the pump decreases from $|\alpha|= 43.9 ^\circ$ to $|\alpha|= 40.8 ^\circ$.
In the mixed case, the possible values of TAM projection for TH light are hence $m^{3\omega} \in \{ 3, 1, -1,-3\}$. 
In our setup, components other than $m^{3\omega}=3$ can only appear if the input pump is not a pure RCP beam. 
It is important to mention that TH generation from a normally-incident CP plane-wave pump is forbidden by selection rules~\cite{Tang1971-SelectionRulesforC} in a film  of isotropic material such as a-Si.

We highlight that the present isotropic case is the simplest one. 
In a general sample, both its shape and its material can have different degrees of rotational symmetry. 
A simple rule has been put forward to deal with the general case~\cite{Nikitina2024-AchiralNanostructure}. 
When a pump beam with $m^{\omega}$ generates the $q$-th harmonic, its final TAM projection $m^{q\omega}$ may be equal to any of the following values:
\begin{align}
    \label{eq:selrul}
    m^{q\omega} = q m^{\omega} + m^\text{tens} + \nu \mathfrak{n}, \ \  \nu \in \mathbb{Z},
\end{align}
where the possible values of $m^\text{tens}$ are determined by the rotational properties of the nonlinear susceptibility tensor $\hat\chi$ of the material, and $\mathfrak{n}$ is the order of the rotational axis of the shape of the nanostructure, that is, the shape is invariant under rotations by angle $2\pi/\mathfrak{n}$. 
In our experiment, $q=3$, and the full rotational symmetry implies $m^\text{tens}=0$, and $\mathfrak{n}\to\infty$, $\nu=0$ in Eq.~(\ref{eq:selrul}). 
When $m^\omega=1$, we recover the tripling case $m^{3\omega}=3$. 

{
The experimental results can be understood by considering the expansion of the pump and TH beams into Bessel beams with well-defined TAM $z$-projection, and well-defined helicity.
Helicity is the generalization of the circular polarization handedness of plane waves onto general Maxwell fields. Any given electromagnetic field $(\mathbf{E},\mathbf{B})$ can be decomposed into its two helicity components $\pm$1 as~\cite{Lakhtakia1994,fernandez2013electromagnetic,tischler2014experimental}:
\begin{align}
\label{eq:Lambda}
    \mathbf{\Lambda}_\pm = \sqrt{\frac{\epsilon}{2}}(\mathbf{E} \pm \iu c\mathbf{B} ),
\end{align}
where $\epsilon$ is the permittivity and $c$ the light velocity in the medium. A field of well-defined helicity is one where all its composing plane waves have the same polarization handedness.
In such a case, only one of the two helical fields in Eq.~\eqref{eq:Lambda} is different than zero. 
The split written in Eq.~\eqref{eq:Lambda} is valid for general Maxwell fields, in particular, collimated fields and focused fields.

Any electromagnetic beam can be expanded into Bessel beams, but they are particularly useful for analyzing cylindrically symmetric experiments \cite{FerCor2012b,tischler2014experimental,FerCorTHESIS} such as the current one. 
This is because both the pump and the TH beams contain very few kinds of Bessel beams with fixed TAM $z$-projection $m$, and helicity (handedness) $\lambda=\pm 1$, which we will denote by $(m,\lambda)$:
\begin{equation}
	\label{eq:bbexp}
	(m,\lambda) {\coloneqq} \int_0^{\theta_{\text{max}}} 
        \dd\theta\sin\theta\ c_{m\lambda}(\theta)\mathcal{B}_{m\lambda}^{k \theta}(\mathbf{r}),
\end{equation}
where $\theta=\acos(k_z/k)$ is the angle of the cone that defines the Bessel beam in Fourier space, $k=\omega/c$,  and $k_z$ is the projection of the wavevector on $z$-axis, which is the same for all the plane waves in such cone, and we assume here that $k_z>0$.
Intuitively, collimated beams only contain very small values of $\theta$ in their expansion ($\theta_{\text{max}} \rightarrow 0$), while focused beams feature larger $\theta_{\text{max}}$. 
The full expression of the Bessel beams $\mathcal{B}_{m\lambda}^{k \theta}(\mathbf{r})$ can be found in Suppl. Inf.~\ref{app:BBs}.
The $c_{m\lambda}(\theta)$ in Eq.~\eqref{eq:bbexp} are complex coefficients, whose explicit value is not necessary to qualitatively explain the experimentally obtained images. 
For a given $(m,\lambda)$, the phase singularities in each polarization are determined by $m$, and $\lambda$ determines which polarization dominates in the collimated limit, as shown in Tab.~\ref{tab:BB}. 
We note that the definition of a right circularly polarized (RCP, unit vector $\rhat$) and left circularly polarized (LCP, unit vector $\lhat$) field used here is opposite to the one used in \cite{FerCor2012b,tischler2014experimental,FerCorTHESIS}. 
Table~\ref{tab:BB} highlights the important fact that in-plane polarization and helicity are not the same thing. 
Beams of well-defined helicity can contain the three polarization components $\vb e_R$, $\vb e_L$, and $\vb e_z$. 
Such situation is rather common. 
For example, a collimated circularly polarized Gaussian beam also has non-zero weights in all three components, but one of them overwhelmingly dominates over the other two. 
Such dominance is reduced upon focusing, which increases $\theta_{\text{max}}$, and with it, the relative amplitudes of the other two components (see Eq.~\eqref{eq:BBs}).
Even a circularly polarized plane wave, which can be seen as the exceptional case of containing a single polarization vector, will acquire some intensity on the other two polarizations upon focusing (see Suppl. Inf.~\ref{sec:foc_pw_an}).

In our case, the imperfection of the QWP implies that we always need to consider the superposition of two Gaussian beams with opposite polarization handedness. }
By matching their respective dominant polarizations, and requiring the presence of light in the center, Tab.~\ref{tab:BB} readily determines that the RCP input pump beam is of kind $(m=1,\lambda=1)$, and the LCP input pump beam of kind $(m=-1,\lambda=-1)$. 

The rotational symmetry will preserve the TAM $z$-projection in the sense explained above. 
The helicity, however, is allowed to change as long as the interaction is not symmetric under the electromagnetic duality transformation~\cite{fernandez2013electromagnetic,Zwanziger1968,Deser1976,Drummond1999}. 
Duality symmetry requires that the electric and magnetic responses of the medium are of the same order, which does not occur in natural materials at optical frequencies. 
In non-dual symmetric systems, the efficiency of helicity conversion can vary~\cite{tischler2014experimental}. 
In order to observe this, we have also conducted a polarization resolved experiment, probing the RCP and LCP of the TH signal; this has been accomplished by introducing a QWP and a polarizer on the TH beam optical path before the imaging camera (see the setup in Suppl. Inf.~\ref{app:setup}). 
Fig.~\ref{polar_res}a,b shows images of the TH signal obtained with pump polarization close to circular, $\alpha$ = $43.9^\circ$, and deviated to $\alpha$ = $37.9^\circ$. 
We see that when the pump is almost circularly polarized ($\alpha$ = $43.9^\circ$), the TH signal is detected only when probing the signal with the same handedness as the pump, while it is not visible when probing the opposite polarization. 
Therefore, we may say that up to the sensitivity of the measurements, helicity is preserved in the TH generated towards the transmission direction in our system. 
Small helicity conversion is also observed in our numerical simulations, as can be appreciated in Fig.~\ref{polar_res}c. 
For our case, we assume, that when three fields at the fundamental frequency combine to produce a TH field $F_1\otimes F_2\otimes F_3\rightarrow T$, the dominant helicity of the TH is determined by the prevailing helicity among the fundamental fields. 

The elliptically polarized pump is composed by two kinds of $(m,\lambda)$ beams: $\approx (1,1)+\varepsilon (-1,-1)$, where $|\varepsilon|$ increases as the ellipticity angle $\alpha$ decreases. 
Then, the allowed TH processes are:
\begin{equation}
	\label{eq:mm}
	\begin{split}
		(1,1)\otimes(1,1)\otimes(1,1)&\rightarrow (3,1)\\
		(1,1)\otimes(1,1)\otimes\varepsilon(-1,-1)&\rightarrow \varepsilon(1,1)\\
		(1,1)\otimes\varepsilon (-1,-1)\otimes\varepsilon(-1,-1)&\rightarrow \varepsilon^2(-1,-1)\\
		\varepsilon (-1,-1)\otimes\varepsilon (-1,-1)\otimes\varepsilon(-1,-1)&\rightarrow \varepsilon^3(-3,-1)\\
	\end{split}
\end{equation}
To 0-th order in $\varepsilon$, we obtain a $(3,1)$ beam, which, according to Tab.~\ref{tab:BB}, features a vortex of charge 2 in its dominant $\rhat$ polarization: $J_{2}\exp(\iu 2\varphi)\rhat$.
This contribution dominates in the top left panel of Fig.~\ref{spiral_exp&sim}, where $\varepsilon\approx 0.02$. 
As $\varepsilon$ grows, other components become more visible ($\varepsilon\approx 0.07$ at $\alpha=40.8^\circ$). 
For example, at first order in $\varepsilon$, we obtain a $(1,1)$ beam: $J_{0}\rhat$, whose intensity maximum is at the center. 
Importantly, for a pure $(3, 1)$ or $(1,1)$ beam, the intensity distribution does not depend on $\varphi$. 
The sequence of images seen when going from the left to the right of the top row of Fig.~\ref{spiral_exp&sim} can be qualitatively understood as the coherent linear superposition of the $(3,1)$ and $(1,1)$ components, where the latter gains relative importance as the ellipticity of the beam grows.
We note that the angular dependence of the generated patterns can be retrieved solely by considering the symmetry of the system (see Suppl. Inf.~\ref{app:theorySiang}).
The instability of optical vortices with charge higher than one \cite{Ricci2012}, which causes their split into single vortices as soon as another beam is present, can be seen from the start of the sequence. The two lobes in the main ring of the intensity patterns also match expectations: The number of lobes equals the difference between the $m^{3\omega}$ of the two interfering beams. 
Finally, the bottom panel of the polarization resolved images in Fig.~\ref{polar_res}b shows a weak TH LCP component with maximum intensity in the center generated by an elliptical beam with $\alpha=37.9$ degrees. 
This is consistent with the $(-1,-1)$ dominant component which appears at order $\varepsilon^2$, which is a beam whose dominant polarization has a maximum at the center and exhibits no phase singularity.
Order $\varepsilon^2$ is also the lowest order at which a component of changed helicity appears.
{The sensitivity of the generated TH signal to the purity of excitation polarization can be utilized for CP light detection~\cite{Hu2024-ReviewofPolarizedL}.}

\subsection{Numerical simulation}
\label{sec:simulation}

\begin{figure*}[t]
\centering
\includegraphics[width=\textwidth]{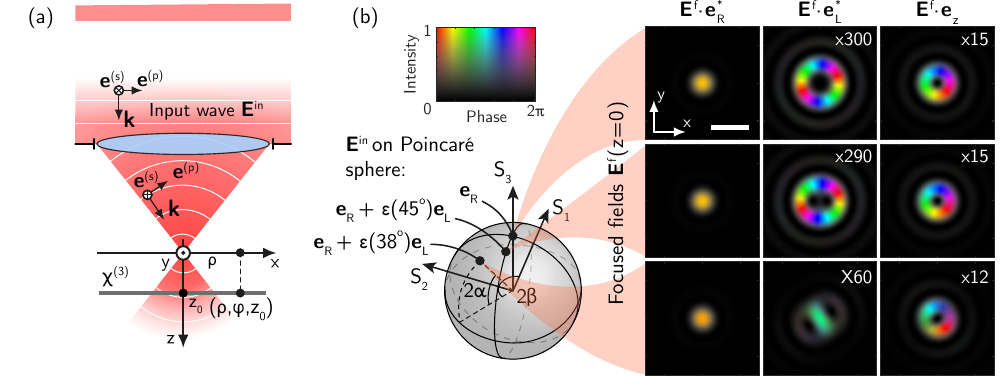}
    \caption{
    a) Schematic of the focusing system and the reference frame. 
    b) Numerically calculated intensity, $|\vb E^{f}\cdot \vb e_i|^2$,  and phase, $\arg(\vb E^{f}\cdot \vb e_i)$, distributions of electric fields in the focal plane $(z_0 = 0)$ of the lens (NA = 0.85) for  input wave $\vb{E}^{\text{in}}$ ellipticity angles $\alpha = 45^\circ, 43.9^\circ$ and $37.9^\circ$,
    from top to bottom. An input RCP plane wave (located at the pole of the Poincar\'e sphere, $\alpha = 45^\circ$) undergoes transformation under focusing, resulting in a beam that includes all three polarization components: $\vb e_R, \vb e_L$ and $\vb e_z$.
    Patterns are normalized to the maximum intensity of the corresponding $|\vb E^{f}\cdot \vb e_R^*|^2$ distribution. Scale bar, 2 $\mu$m. }
\label{pump_sch}
\end{figure*}

In our experiment, the size of the entrance pupil of the objective focusing the pump laser was much smaller than the diameter of the incident Gaussian beams, giving nearly uniform illumination of the entrance aperture. Therefore, we can approximate the pump beam entering the objective as a linear combination of circularly polarized plane waves:
\begin{equation} 
    \label{eq:inp}
	\vb{E}^{\text{in}}\propto\rhat+\varepsilon(\gamma)\lhat.
\end{equation} 
{where the parameter $\varepsilon(\gamma)$ determines the contribution of the LCP wave and can be expressed through the angle $\gamma$ between the direction of oscillation of the input $y$-polarized light and the fast axis of the wave plate, and its retardance angle $\tau$ (see Suppl. Inf.~\ref{app:vareps_circ}).}
The analytical expression of the focused electric field after the lens can be obtained (see Eq. \ref{eq:foc_orig})~\cite{richards1959electromagnetic, Kant1993-AnAnalyticalSolutio}. 
Figure~\ref{pump_sch}a illustrates the reference frame. 
In terms of $s$- and $p$-polarized plane waves, with polarization vectors determined as follows~\cite{novotny2012principles}:
\begin{align}
    \mathbf{e}^{\text{(s)}} = \frac{1}{k_{\rho}}[k_y, -k_x, 0], \\
    \mathbf{e}^{\text{(p)}} = \frac{1}{k_{\rho}k}[k_zk_x, k_zk_y, -k_{\rho}^2],
\end{align}
the focused electric field is given by:
\begin{align}
    \nonumber 
        \mathbf{E}^\text{f}(\rho,\varphi,z) =A\int\limits_{k_x^2+k_y^2<k_{\rho0}^2} \frac{g(k_z)}{k_z}
    [(\mathbf{E}^{\text{in}}\cdot \mathbf{e}^{\text{(s)}})\mathbf{e}^\text{(s)} + \\
	+(\mathbf{E}^{\text{in}}\cdot (\vb{e}_z \times \mathbf{e}^\text{(s)}))\mathbf{e}^\text{(p)} ] e^{\iu\mathbf{k\cdot r}}
    \dd k_x\dd k_y,
    \label{foc_pw}
\end{align}
where $A$ is the normalization constant, $g(k_z)$ is the apodization factor ($g(k_z)=\sqrt{k_z/k}$ for an aplanatic focusing system~\cite{richards1959electromagnetic, Kant1993-AnAnalyticalSolutio}), $k=\omega/c$, $k_{\rho0}$ is the maximum value of the in-plane projection of the wave vector and it is equal to $k\mathrm{NA}$, $k_q$ (with $q=x,y,z$) are the Cartesian components of the wavevector, and $\mathbf{E}^{\text{in}}$ is the electric field of the input plane wave. The field in the form of an integral in Eq.~\eqref{foc_pw} can be retrieved using efficient computational algorithms~\cite{hu2020efficient}. 

Figure~\ref{pump_sch}b shows the intensity and phase distributions of focused plane waves with $\alpha = 45^\circ, 43.9^\circ$ and $37.9^\circ$ projected onto $\rhat$, $\lhat$ and $\zhat$ basis vectors.
For a pure RCP plane wave ($\alpha = 45^\circ$), the intensity of the electric field component with the same polarization as the input, $\vb E^{f}\cdot \vb e_R^*$, is maximal and its phase does not exhibit a helical profile. 
A helical phase is present for the oppositely polarized, $\vb E^{f}\cdot \vb e_L^*$, and longitudinal components, $\vb E^{f}\cdot \vb e_z$, as expected. 
The input RCP plane wave can be seen as a $(1,1)$ beam in Tab.~\ref{tab:BB} with $\theta_{\text{max}}\rightarrow0$, where the only significant polarization component is the one which features a $J_0$ Bessel function without a phase singularity. 
The other two polarizations feature Bessel functions of higher order, whose amplitude is much suppressed by their vanishing $k\sin\theta\rho$ argument, and the $\theta$-dependent factors multiplying the Bessel functions. 
Upon focusing, which preserves both $m$ and $\lambda$ \cite{fernandez2013electromagnetic}, the value of $\theta_{\text{max}}$ increases, and with it the relative strength of the other two polarizations, with their corresponding phase singularities: $J_2\exp(\ii2\varphi)\lhat$, and $J_1\exp(\ii\varphi)\zhat$.

{When using an imperfect quarter-wave plate, the maximum absolute value of ellipticity angle is achieved at $\gamma = 45^\circ$, and, in our case, is equal to $\alpha = 43.9^\circ$. 
Therefore, the point representing the polarization state of the input radiation shifts from the pole of the Poincar\'e sphere. At the same time, the oppositely polarized component of the focused field, $\vb E^{f}\cdot \vb e_L^*$, exhibits some asymmetry and an increase in intensity (see Fig.~\ref{pump_sch}b, middle panel). Further reduction of the ellipticity angle, achieved by deviating the wave plate to $\gamma = 38^\circ$ ($\alpha = 37.9^\circ$), leads to a further increase in the intensity of $\vb E^{f}\cdot \vb e_L^*$ component and a significant change in its distribution, with a maximum emerging at the center (see Fig.~\ref{pump_sch}b, bottom panel).}

In the numerical model, we approximate the a-Si film by an infinitely thin nonlinear layer placed in air at $z=z_0$. The nonlinear polarization density for the TH generation in the nonlinear layer is written as \cite{Boyd2003}:
\begin{equation}
    {\vb P^{3\omega}}(\rho, \varphi) = \hat\chi : (\mathbf{E}^{\omega}(\rho, \varphi) \otimes \mathbf{E}^{\omega}(\rho, \varphi) \otimes \mathbf{E}^{\omega}(\rho, \varphi)),
    \label{polar_3}
\end{equation}
where 
$\mathbf{E}^{\omega}(\rho, \varphi)=\mathbf{E}^\text{f}(\rho, \varphi, z_0)$ is determined from Eq.~\eqref{foc_pw}. 
The components of $\hat\chi$ can be found in Suppl. Inf.~\ref{app:chi3}.  
The nonlinear polarization source generates electric fields that are calculated using the Green's function approach~\cite{Sipe:87}. 

We find that misalignment along the propagation axis between the pump beam focus position and the film plane affects the shape of the TH field. In order to achieve the best match between the simulation and the experimental data we introduce a shift in the position of the waist of the pump beam by $z_0 = -2.055 \ \mu$m. Figure \ref{spiral_exp&sim}b shows the TH patterns calculated  numerically. 
In the model we neglected the presence of the substrate, because it does not qualitatively affect the results (see Suppl. Inf.~\ref{app:sim_subs}). 
The numerically calculated patterns were averaged over the focal depth of the collecting objective lens (NA = 0.4). 

\subsection{Helicity change in the TH simulation}

Since focusing preserves $\lambda$, the pump electric field produced by a focused RCP plane wave can be represented as a sum of plane waves of the same handedness, by construction, giving the ratio $\int|\mathbf{\Lambda_-}|^2\dd S/\int|\mathbf{\Lambda_+}|^2\dd S = 0$, where the surface integrals are performed in the plane $z=z_0$. 
In our model, the duality symmetry inherent in vacuum is broken by the non-linear interaction in an obvious way: The absence of magnetic terms in Eq.~\eqref{polar_3} negates the electric-magnetic equivalence required by duality.
For the TH generated by a focused RCP plane wave we obtain numerically $\int|\mathbf{\Lambda_-}|^2\dd S/\int|\mathbf{\Lambda_+}|^2 \dd S = 1.43\times10^{-4}$, which implies a rather small helicity conversion factor. 

Figure~\ref{polar_res}c shows the helicity of a pump beam with ellipticity angle $\alpha = 43.9^{\circ}$ and generated TH signal. 
In this case, for the pump $\int|\mathbf{\Lambda_-}|^2\dd S/\int|\mathbf{\Lambda_+}|^2\dd S = 3.88\times10^{-4}$ becomes nonzero, due to the presence of an opposite polarized component at the input ($\varepsilon \neq 0$ in Eq.~\eqref{eq:inp}). 
Bottom panel of Fig.~\ref{polar_res}c shows the helicity of the corresponding nonlinear signal. 
The change of helicity is again small, and the relative strength $\int|\mathbf{\Lambda_-}|^2\dd S/\int|\mathbf{\Lambda_+}|^2 \dd S= 1.1\times10^{-3}$ prevents one from experimentally detecting a signal in the bottom-left panel of Fig.~\ref{polar_res}a. 

It should be noted that the relatively small helicity changes have been obtained for the TH generated towards the transmission direction, and they do not rule out a much larger helicity conversion in reflection.

{\section{Conclusion}
In this work, we explored light structuring via nonlinear total angular momentum (TAM) addition in thin films of amorphous silicon.
Through both experimental measurements and numerical simulations, we demonstrated the capability of thin layers of isotropic material to enable tripling of the TAM projection under tightly focused laser excitation. 
Our theoretical framework, based on TAM and helicity considerations, provides a powerful and clear approach for analyzing similar experiments and can be applied to the design of new optical devices that exploit symmetry for enhanced control of light structure.
Nonlinear thin layers offer exciting functionalities, combining ultra-compact design with relaxed phase-matching conditions, making them suitable for various applications, including circularly polarized light detection.
This study advances the understanding of nonlinear light-matter interactions using flat isotropic media to achieve sophisticated light tailoring and contributes to the ongoing development of nonlinear flat optics.
}
\section{Methods} 
\label{sec:methods}
A thin layer (1 $\mu$m) of amorphous silicon was deposited by means of electron beam (e-beam) evaporation (rate 0.4 nm/s, base vacuum $2\times10^{-7}$ mbar) on a fused silica (SiO$_2$) substrate previously cleaned using acidic piranha solution (H$_2$SO$_4$:H$_2$O$_2$).

As a pump laser source we used the emission of an optical parametric amplifier (Coherent Opera-F) coupled with a femtosecond laser system (Coherent MONACO), operating at 1035~nm with the pulse duration of $\approx$300~fs and 500~kHz repetition rate. 
The pump light at wavelength of 1500~nm was focused on the sample with an Olympus LCPlan N 100X/0.85 IR objective, the TH signal was collected by an Olympus LMPlanFL N 20x/0.40 objective and filtered by a short pass filter (Thorlabs FESH0950). 
To transform the polarization of a linearly polarized input light, we used an achromatic quarter-wave plate (Thorlabs AQWP10M-1600) with a retardance of 0.2437 at 1500~nm. 
To inspect the polarization of the generated TH, we specifically employed a quarter-wave plate (Thorlabs AQWP10M-580) and a polarizer (Thorlabs GL10-A).

\section{Acknowledgements}
This work was partially supported by the European Union under the Italian National Recovery and Resilience Plan (NRRP) of NextGenerationEU, of partnership on “Telecommunications of the Future” (PE00000001 - program “RESTART”), S2 SUPER – Programmable Networks,  Cascade project PRISM - CUP: E13C22001870001, 
PRIN 2020 project METEOR (2020EY2LJT), METAFAST project that received funding
from the European Union Horizon 2020 Research and Innovation programme
under Grant Agreement No. 899673.

\section{Author contribution statement}
Conceptualization: EM, DdC, CDA. Data curation: PF, AT, EM. Formal analysis: PF, KF, IFK, AT, EM. Funding: CDA, AT, ACC. Investigation: EM, AT, PF. Methodology: IFC, KF. Project administration: CDA, DdC. Resources: DG, CDA, MP, DdC. Software: EM, DdC. Supervision: DdC, KF, MP, CDA, ACC. Visualization: EM. Writing original draft: EM, KF, ICF, DdC, CDA. Writing, review and editing: all authors. 


\normalem
\bibliographystyle{unsrturl}  

\bibliography{sample_B} 

\begin{thebibliography}{10}

\bibitem{Forbes2021-Structuredlight}
Andrew Forbes, Michael de~Oliveira, and Mark~R. Dennis.
\newblock {Structured light}.
\newblock {\em Nat. Photonics}, 15(4):253--262, 2021.
\newblock \href {https://doi.org/10.1038/s41566-021-00780-4}
  {\path{doi:10.1038/s41566-021-00780-4}}.

\bibitem{Davis:00}
Jeffrey~A. Davis, Dylan~E. McNamara, Don~M. Cottrell, and Juan Campos.
\newblock {Image processing with the radial Hilbert
  transform:{\hspace{1em}}theory and experiments}.
\newblock {\em Opt. Lett.}, 25(2):99--101, 2000.
\newblock \href {https://doi.org/10.1364/OL.25.000099}
  {\path{doi:10.1364/OL.25.000099}}.

\bibitem{furhapter2005spiral}
Severin F{\ifmmode\ddot{u}\else\"{u}\fi}rhapter, Alexander Jesacher, Stefan
  Bernet, and Monika Ritsch-Marte.
\newblock {Spiral phase contrast imaging in microscopy}.
\newblock {\em Opt. Express}, 13(3):689--694, 2005.
\newblock URL: \url{https://doi.org/10.1364/opex.13.000689}.

\bibitem{klar2001breaking}
T.~A. Klar, E.~Engel, and S.~W. Hell.
\newblock {Breaking Abbe's diffraction resolution limit in fluorescence
  microscopy with stimulated emission depletion beams of various shapes}.
\newblock {\em Phys. Rev. E}, 64(6):Pt, 2001.
\newblock \href {https://doi.org/10.1103/PhysRevE.64.066613}
  {\path{doi:10.1103/PhysRevE.64.066613}}.

\bibitem{kozawa2018superresolution}
Yuichi Kozawa, Daichi Matsunaga, and Shunichi Sato.
\newblock {Superresolution imaging via superoscillation focusing of a radially
  polarized beam}.
\newblock {\em Optica}, 5(2):86--92, 2018.
\newblock \href {https://doi.org/10.1364/OPTICA.5.000086}
  {\path{doi:10.1364/OPTICA.5.000086}}.

\bibitem{Yoshida:19}
Mio Yoshida, Yuichi Kozawa, and Shunichi Sato.
\newblock {Subtraction imaging by the combination of higher-order vector beams
  for enhanced spatial resolution}.
\newblock {\em Opt. Lett.}, 44(4):883--886, 2019.
\newblock \href {https://doi.org/10.1364/OL.44.000883}
  {\path{doi:10.1364/OL.44.000883}}.

\bibitem{d2013photonic}
Vincenzo D'Ambrosio, Nicol{\ifmmode\grave{o}\else\`{o}\fi} Spagnolo, Lorenzo
  Del~Re, Sergei Slussarenko, Ying Li, Leong~Chuan Kwek, Lorenzo Marrucci,
  Stephen~P. Walborn, Leandro Aolita, and Fabio Sciarrino.
\newblock {Photonic polarization gears for ultra-sensitive angular
  measurements}.
\newblock {\em Nat. Commun.}, 4(2432):1--8, 2013.
\newblock \href {https://doi.org/10.1038/ncomms3432}
  {\path{doi:10.1038/ncomms3432}}.

\bibitem{wang2012terabit}
Jian Wang, Jeng-Yuan Yang, Irfan~M. Fazal, Nisar Ahmed, Yan Yan, Hao Huang,
  Yongxiong Ren, Yang Yue, Samuel Dolinar, Moshe Tur, and Alan~E. Willner.
\newblock {Terabit free-space data transmission employing orbital angular
  momentum multiplexing}.
\newblock {\em Nat. Photonics}, 6(7):488--496, 2012.
\newblock \href {https://doi.org/10.1038/nphoton.2012.138}
  {\path{doi:10.1038/nphoton.2012.138}}.

\bibitem{vallone2014free}
Giuseppe Vallone, Vincenzo D{'}Ambrosio, Anna Sponselli, Sergei Slussarenko,
  Lorenzo Marrucci, Fabio Sciarrino, and Paolo Villoresi.
\newblock {Free-Space Quantum Key Distribution by Rotation-Invariant Twisted
  Photons}.
\newblock {\em Phys. Rev. Lett.}, 113(6):060503, 2014.
\newblock \href {https://doi.org/10.1103/PhysRevLett.113.060503}
  {\path{doi:10.1103/PhysRevLett.113.060503}}.

\bibitem{nicolas2014quantum}
A.~Nicolas, L.~Veissier, L.~Giner, E.~Giacobino, D.~Maxein, and J.~Laurat.
\newblock {A quantum memory for orbital angular momentum photonic qubits}.
\newblock {\em Nat. Photonics}, 8(3):234--238, 2014.
\newblock \href {https://doi.org/10.1038/nphoton.2013.355}
  {\path{doi:10.1038/nphoton.2013.355}}.

\bibitem{garcia2024topological}
Ana Garc{\ifmmode\acute{\imath}\else\'{\i}\fi}a-Cabrera, Roberto
  Boyero-Garc{\ifmmode\acute{\imath}\else\'{\i}\fi}a,
  {\ifmmode\acute{O}\else\'{O}\fi}scar
  Zurr{\ifmmode\acute{o}\else\'{o}\fi}n-Cifuentes, Javier Serrano, Julio~San
  Rom{\ifmmode\acute{a}\else\'{a}\fi}n, Luis Plaja, and Carlos
  Hern{\ifmmode\acute{a}\else\'{a}\fi}ndez-Garc{\ifmmode\acute{\imath}\else\'{\i}\fi}a.
\newblock {Topological high-harmonic spectroscopy}.
\newblock {\em Commun. Phys.}, 7(28):1--10, 2024.
\newblock \href {https://doi.org/10.1038/s42005-023-01511-7}
  {\path{doi:10.1038/s42005-023-01511-7}}.

\bibitem{padgett2011tweezers}
Miles Padgett and Richard Bowman.
\newblock {Tweezers with a twist}.
\newblock {\em Nat. Photonics}, 5(6):343--348, 2011.
\newblock \href {https://doi.org/10.1038/nphoton.2011.81}
  {\path{doi:10.1038/nphoton.2011.81}}.

\bibitem{Lin2018-All-opticalmachinel}
Xing Lin, Yair Rivenson, Nezih~T. Yardimci, Muhammed Veli, Yi~Luo, Mona
  Jarrahi, and Aydogan Ozcan.
\newblock {All-optical machine learning using diffractive deep neural
  networks}.
\newblock {\em Science}, 361(6406):1004--1008, 2018.
\newblock \href {https://doi.org/10.1126/science.aat8084}
  {\path{doi:10.1126/science.aat8084}}.

\bibitem{Wang2024-Large-scalephotonic}
Hao Wang, Jianqi Hu, Andrea Morandi, Alfonso Nardi, Fei Xia, Xuanchen Li,
  Romolo Savo, Qiang Liu, Rachel Grange, and Sylvain Gigan.
\newblock {Large-scale photonic computing with nonlinear disordered media}.
\newblock {\em Nat. Comput. Sci.}, 4(6):429--439, 2024.
\newblock \href {https://doi.org/10.1038/s43588-024-00644-1}
  {\path{doi:10.1038/s43588-024-00644-1}}.

\bibitem{devlin2017}
Robert~C. Devlin, Antonio Ambrosio, Noah~A. Rubin, J.~P.~Balthasar Mueller, and
  Federico Capasso.
\newblock {Arbitrary spin-to{\textendash}orbital angular momentum conversion of
  light}.
\newblock {\em Science}, 358(6365):896--901, 2017.
\newblock \href {https://doi.org/10.1126/science.aao5392}
  {\path{doi:10.1126/science.aao5392}}.

\bibitem{dorrah2022}
Ahmed~H. Dorrah and Federico Capasso.
\newblock {Tunable structured light with flat optics}.
\newblock {\em Science}, 376(6591), 2022.
\newblock \href {https://doi.org/10.1126/science.abi6860}
  {\path{doi:10.1126/science.abi6860}}.

\bibitem{Akhiezer}
A.~I. Akhiezer and V.~B. Berestetsky.
\newblock {\em {Quantum Electrodynamics}}.
\newblock Interscience Publishers, Hoboken, NJ, USA, 1965.

\bibitem{FerCor2012b}
Ivan Fernandez-Corbaton, Xavier Zambrana-Puyalto, and Gabriel Molina-Terriza.
\newblock {Helicity and angular momentum: A symmetry-based framework for the
  study of light-matter interactions}.
\newblock {\em Phys. Rev. A}, 86(4):042103, 2012.
\newblock \href {https://doi.org/10.1103/PhysRevA.86.042103}
  {\path{doi:10.1103/PhysRevA.86.042103}}.

\bibitem{tischler2014experimental}
Nora Tischler, Ivan Fernandez-Corbaton, Xavier Zambrana-Puyalto, Alexander
  Minovich, Xavier Vidal, Mathieu~L. Juan, and Gabriel Molina-Terriza.
\newblock {Experimental control of optical helicity in nanophotonics}.
\newblock {\em Light Sci. Appl.}, 3(6):e183, 2014.
\newblock \href {https://doi.org/10.1038/lsa.2014.64}
  {\path{doi:10.1038/lsa.2014.64}}.

\bibitem{Zambrana2016}
Xavier Zambrana-Puyalto, Xavier Vidal, Ivan Fernandez-Corbaton, and Gabriel
  Molina-Terriza.
\newblock {Far-field measurements of vortex beams interacting with nanoholes}.
\newblock {\em Sci. Rep.}, 6(22185):1--10, 2016.
\newblock \href {https://doi.org/10.1038/srep22185}
  {\path{doi:10.1038/srep22185}}.

\bibitem{Zwanziger1968}
Daniel Zwanziger.
\newblock {Quantum Field Theory of Particles with Both Electric and Magnetic
  Charges}.
\newblock {\em Phys. Rev.}, 176(5):1489--1495, 1968.
\newblock \href {https://doi.org/10.1103/PhysRev.176.1489}
  {\path{doi:10.1103/PhysRev.176.1489}}.

\bibitem{Weinberg2005-TheQuantumTheoryof}
Steven Weinberg.
\newblock {\em {The Quantum Theory of Fields}}.
\newblock Cambridge University Press, Cambridge, England, UK, 1995.
\newblock \href {https://doi.org/10.1017/CBO9781139644167}
  {\path{doi:10.1017/CBO9781139644167}}.

\bibitem{Tung2020-GroupTheoryInPhysi}
Wu-Ki Tung.
\newblock {\em {Group Theory in Physics}}.
\newblock World Scientific Publishing Company, Singapore, 1985.
\newblock \href {https://doi.org/10.1142/0097} {\path{doi:10.1142/0097}}.

\bibitem{Stratton}
Julius~Adams Stratton.
\newblock {\em {Electromagnetic Theory}}.
\newblock 2015.
\newblock \href {https://doi.org/10.1002/9781119134640}
  {\path{doi:10.1002/9781119134640}}.

\bibitem{Stein1961-Additiontheoremsfor}
Seymour Stein.
\newblock {Addition theorems for spherical wave functions}.
\newblock {\em Q. Appl. Math.}, 19(1):15--24, 1961.
\newblock \href {https://doi.org/10.1090/qam/120407}
  {\path{doi:10.1090/qam/120407}}.

\bibitem{Hacyan2006}
S.~Hacyan and R.~J{\ifmmode\acute{a}\else\'{a}\fi}uregui.
\newblock {A relativistic study of Bessel beams}.
\newblock {\em J. Phys. B: At. Mol. Opt. Phys.}, 39(7):1669, 2006.
\newblock \href {https://doi.org/10.1088/0953-4075/39/7/009}
  {\path{doi:10.1088/0953-4075/39/7/009}}.

\bibitem{Afanasev2013}
Andrei Afanasev, Carl~E. Carlson, and Asmita Mukherjee.
\newblock {Off-axis excitation of hydrogenlike atoms by twisted photons}.
\newblock {\em Phys. Rev. A}, 88(3):033841, 2013.
\newblock \href {https://doi.org/10.1103/PhysRevA.88.033841}
  {\path{doi:10.1103/PhysRevA.88.033841}}.

\bibitem{ikeda2007enhanced}
Kazuhiro Ikeda, Yaoming Shen, and Yeshaiahu Fainman.
\newblock {Enhanced optical nonlinearity in amorphous silicon and its
  application to waveguide devices}.
\newblock {\em Opt. Express}, 15(26):17761--17771, 2007.
\newblock \href {https://doi.org/10.1364/OE.15.017761}
  {\path{doi:10.1364/OE.15.017761}}.

\bibitem{Boyd2003}
Robert Boyd.
\newblock {\em {Nonlinear Optics}}.
\newblock Academic Press, Cambridge, MA, USA, Dec 2002.
\newblock URL:
  \url{https://www.elsevier.com/books/nonlinear-optics/boyd/978-0-12-121682-5}.

\bibitem{collett1992polarized}
Edward Collett.
\newblock Polarized light. fundamentals and applications.
\newblock {\em Optical Engineering}, 1992.

\bibitem{Freter2024}
Lukas Freter, Benedikt Zerulla, Marjan Krsti{\ifmmode\acute{c}\else\'{c}\fi},
  Christof Holzer, Carsten Rockstuhl, and Ivan Fernandez-Corbaton.
\newblock {Tensor product space for studying the interaction of bipartite
  states of light with nanostructures}.
\newblock {\em Phys. Rev. A}, 110(4):043516, 2024.
\newblock \href {https://doi.org/10.1103/PhysRevA.110.043516}
  {\path{doi:10.1103/PhysRevA.110.043516}}.

\bibitem{Tang1971-SelectionRulesforC}
C.~L. Tang and Herbert Rabin.
\newblock {Selection Rules for Circularly Polarized Waves in Nonlinear Optics}.
\newblock {\em Phys. Rev. B}, 3(12):4025--4034, 1971.
\newblock \href {https://doi.org/10.1103/PhysRevB.3.4025}
  {\path{doi:10.1103/PhysRevB.3.4025}}.

\bibitem{Nikitina2024-AchiralNanostructure}
Anastasia Nikitina and Kristina Frizyuk.
\newblock {Achiral Nanostructures: Perturbative Harmonic Generation and
  Dichroism Under Vortex and Vector Beams Illumination}.
\newblock {\em Adv. Opt. Mater.}, 12(25):2400732, 2024.
\newblock \href {https://doi.org/10.1002/adom.202400732}
  {\path{doi:10.1002/adom.202400732}}.

\bibitem{Lakhtakia1994}
A.~Lakhtakia.
\newblock {\em {Beltrami Fields in Chiral Media {$\vert$} World Scientific
  Series in Contemporary Chemical Physics}}, volume~2.
\newblock World Scientific Publishing Company, Singapore, 1994.
\newblock \href {https://doi.org/10.1142/2031} {\path{doi:10.1142/2031}}.

\bibitem{fernandez2013electromagnetic}
Ivan Fernandez-Corbaton, Xavier Zambrana-Puyalto, Nora Tischler, Xavier Vidal,
  Mathieu~L. Juan, and Gabriel Molina-Terriza.
\newblock {Electromagnetic Duality Symmetry and Helicity Conservation for the
  Macroscopic Maxwell's Equations}.
\newblock {\em Phys. Rev. Lett.}, 111(6):060401, 2013.
\newblock \href {https://doi.org/10.1103/PhysRevLett.111.060401}
  {\path{doi:10.1103/PhysRevLett.111.060401}}.

\bibitem{FerCorTHESIS}
Ivan Fernandez-Corbaton.
\newblock {\em Helicity and duality symmetry in light matter interactions:
  Theory and applications}.
\newblock PhD thesis, Macquarie University, 2014.
\newblock {arXiv}: 1407.4432.
\newblock \href {https://doi.org/10.48550/arXiv.1407.4432}
  {\path{doi:10.48550/arXiv.1407.4432}}.

\bibitem{Deser1976}
Stanley Deser and Claudio Teitelboim.
\newblock {Duality transformations of Abelian and non-Abelian gauge fields}.
\newblock {\em Phys. Rev. D}, 13(6):1592--1597, 1976.
\newblock \href {https://doi.org/10.1103/PhysRevD.13.1592}
  {\path{doi:10.1103/PhysRevD.13.1592}}.

\bibitem{Drummond1999}
P.~D. Drummond.
\newblock {Dual symmetric Lagrangians and conservation laws}.
\newblock {\em Phys. Rev. A}, 60(5):R3331--R3334(R), 1999.
\newblock \href {https://doi.org/10.1103/PhysRevA.60.R3331}
  {\path{doi:10.1103/PhysRevA.60.R3331}}.

\bibitem{Ricci2012}
F.~Ricci, W.~L{\ifmmode\ddot{o}\else\"{o}\fi}ffler, and M.~P. van Exter.
\newblock {Instability of higher-order optical vortices analyzed with a
  multi-pinhole interferometer}.
\newblock {\em Opt. Express}, 20(20):22961--22975, 2012.
\newblock \href {https://doi.org/10.1364/OE.20.022961}
  {\path{doi:10.1364/OE.20.022961}}.

\bibitem{Hu2024-ReviewofPolarizedL}
Renjie Hu and Wei Qin.
\newblock {Review of Polarized Light-Spin/Dipole Interactions: Fundamental
  Physics and Application in Circularly Polarized Detecting}.
\newblock {\em Laser Photonics Rev.}, n/a(n/a):2400761, 2024.
\newblock \href {https://doi.org/10.1002/lpor.202400761}
  {\path{doi:10.1002/lpor.202400761}}.

\bibitem{richards1959electromagnetic}
Bernard Richards and Emil Wolf.
\newblock Electromagnetic diffraction in optical systems, ii. structure of the
  image field in an aplanatic system.
\newblock {\em Proceedings of the Royal Society of London. Series A.
  Mathematical and Physical Sciences}, 253(1274):358--379, 1959.
\newblock \href {https://doi.org/10.1098/rspa.1959.0200}
  {\path{doi:10.1098/rspa.1959.0200}}.

\bibitem{Kant1993-AnAnalyticalSolutio}
Rishi Kant.
\newblock {An Analytical Solution of Vector Diffraction for Focusing Optical
  Systems}.
\newblock {\em J. Mod. Opt.}, 1993.
\newblock URL:
  \url{https://www.tandfonline.com/doi/abs/10.1080/09500349314550341}.

\bibitem{novotny2012principles}
Lukas Novotny and Bert Hecht.
\newblock {\em Principles of nano-optics}.
\newblock Cambridge university press, 2012.

\bibitem{hu2020efficient}
Yanlei Hu, Zhongyu Wang, Xuewen Wang, Shengyun Ji, Chenchu Zhang, Jiawen Li,
  Wulin Zhu, Dong Wu, and Jiaru Chu.
\newblock {Efficient full-path optical calculation of scalar and vector
  diffraction using the Bluestein method}.
\newblock {\em Light Sci. Appl.}, 9(119):1--11, 2020.
\newblock \href {https://doi.org/10.1038/s41377-020-00362-z}
  {\path{doi:10.1038/s41377-020-00362-z}}.

\bibitem{Sipe:87}
J.~E. Sipe.
\newblock New green-function formalism for surface optics.
\newblock {\em J. Opt. Soc. Am. B}, 4(4):481--489, Apr 1987.
\newblock \href {https://doi.org/10.1364/JOSAB.4.000481}
  {\path{doi:10.1364/JOSAB.4.000481}}.

\bibitem{Woit2017-QuantumTheoryGroup}
Peter Woit.
\newblock {\em {Quantum Theory, Groups and Representations: An Introduction}}.
\newblock Springer, Berlin, Germany, 2017.
\newblock URL: \url{https://www.math.columbia.edu/~woit/QM/qmbook.pdf}.

\bibitem{Frizyuk2021-NonlinearCircularDi}
Kristina Frizyuk, Elizaveta Melik-Gaykazyan, Jae-Hyuck Choi, Mihail~I. Petrov,
  Hong-Gyu Park, and Yuri Kivshar.
\newblock {Nonlinear Circular Dichroism in Mie-Resonant Nanoparticle Dimers}.
\newblock {\em Nano Lett.}, 21(10):4381--4387, 2021.
\newblock \href {https://doi.org/10.1021/acs.nanolett.1c01025}
  {\path{doi:10.1021/acs.nanolett.1c01025}}.

\bibitem{Shurcliff+1962}
William~A. Shurcliff.
\newblock {\em {Polarized Light}}.
\newblock Harvard University Press, Cambridge, MA, USA, 2013.
\newblock \href {https://doi.org/10.4159/harvard.9780674424135}
  {\path{doi:10.4159/harvard.9780674424135}}.

\bibitem{Hishikawa_1991}
Yoshihiro Hishikawa, Noboru Nakamura, Shinya Tsuda, Shoichi Nakano, Yasuo
  Kishi~Yasuo Kishi, and Yukinori Kuwano~Yukinori Kuwano.
\newblock {Interference-Free Determination of the Optical Absorption
  Coefficient and the Optical Gap of Amorphous Silicon Thin Films}.
\newblock {\em Jpn. J. Appl. Phys.}, 30(5R):1008, 1991.
\newblock \href {https://doi.org/10.1143/JJAP.30.1008}
  {\path{doi:10.1143/JJAP.30.1008}}.

\end{thebibliography}

\clearpage
\newpage
\appendix
\section*{Supplementary Information}

\setcounter{equation}{0}\renewcommand\theequation{S\arabic{equation}}
\setcounter{figure}{0}\renewcommand\thefigure{S\arabic{figure}}

\section{Definitions\label{sec:definitions}}
\textbf{Total angular momentum of a field} is a number \(J\in\mathbb{N}_0 \), labeling the irreducible representation of the group \( \mathrm{SO}(3) \) with dimension \( 2J + 1 \) and characterizing the field’s behavior under rotations in 3D space \cite[11.4.2]{Tung2020-GroupTheoryInPhysi}, \cite[``Eigenfunctions of the Photon Angular Momentum Operator'']{Akhiezer}.

There exist \( 2J + 1 \) eigenfunctions of the total angular momentum operator \( \hat{J}^2 \) with eigenvalue $J(J+1)$, which transform through each other according to the \( J \)-th representation of the rotation group SO(3). 
One of the simplest examples are spherical functions $Y_{Jm}$ \cite[8.4]{Woit2017-QuantumTheoryGroup}. 
Another examples are electric or magnetic vector spherical harmonics.

\textbf{Total angular momentum projection of a field} is a number \(m\in\mathbb{Z} \), characterizing the field’s behavior under rotations around the $z$-axis.
The function characterized by TAM projection $m$  does not change its shape and acquires a phase \( e^{\iu m \alpha} \) under such rotations, where \( \alpha \) is the angle of rotation.

Note that one can also define the TAM projection through the representations of the symmetry group of rotations of a cone (all rotations around a single axis), and the field can possess a well-defined $ m $ even if it consists of a sum of all possible $ J \geq |m| $.

\textbf{Helicity of a field}

As an operator,  helicity $\Lambda$ is the projection of the angular momentum vector $\mathbf{J}$ onto the direction of the linear momentum vector $\mathbf{P}$,
\begin{equation}
	\label{eq:heldef}
	\Lambda=\frac{\mathbf{J}\cdot\mathbf{P}}{|\mathbf{P}|}.
\end{equation}
Equation~(\ref{eq:heldef}) is the most general form of the helicity operator, which is valid for many particles and fields, such as electrons and gravitational waves. For Maxwell fields, helicity describes the sense of screw in light: The circular polarization handedness.

Helicity characterizes the behavior of electromagnetic fields under duality transformations. Fields with helicity $\lambda$ just acquire a phase \( e^{-\iu \lambda \vartheta} \) under such transformations, where \( \vartheta \) is the ``angle'', characterizing the duality transformation~\cite{FerCorTHESIS}.

In the context of representation theory, the eigenvalues of the helicity operator, $\lambda=\pm 1$ in case of photons, label the irreps of the little group of the Poincar\'e group E(2), for massless particles, \cite[10.4.4]{Tung2020-GroupTheoryInPhysi}.

\section{Bessel Beams of well-defined helicity\label{app:BBs}}
A Bessel beam can be constructed as a weighted integral sum of all the plane waves with the same frequency $\omega$, $z$-component of the wavevector $k_z$, and polarization.
Different polarization basis can be used. 
When using the linear polarization basis the plane waves can be selected to be either all $s$-polarized or all $p$-polarized~\cite{Hacyan2006}, whereas when circular polarization handedness are used~\cite{Afanasev2013,FerCor2012b}, all the plane waves have either positive or negative helicity. 
The wavevectors of all the plane waves in the integral determine a conical surface in $k$-space, the angle of the cone is equal to $\theta=\acos\left(k_z/k\right)$. 
The weights in the integral are selected so that, upon rotation around the $z$ axis, the beam maintains its form and picks up a phase equal to $\exp(-\ii m\varphi)$. 
That is, the Bessel beams are eigenstates of the TAM projection operator with eigenvalue $m$. 
If the handedness basis is chosen for the plane waves, they are also eigenstates of the helicity operator with eigenvalues $\lambda=1$ or $\lambda=-1$. 
For a fixed wavenumber $k=\omega/c$, the explicit forms of the Bessel beams $\mathcal{B}_{m\lambda}^{k \theta}(\mathbf{r})$ in cylindrical coordinates $[\rho=\sqrt{x^2+y^2},\varphi=\atan(y,x),z]$ are \cite[Eq.~(2.82)]{FerCorTHESIS}, \cite[Eq.~(11)]{Afanasev2013}:
\begin{equation}
	\begin{split}
	\label{eq:BBs}
		&\mathcal{B}_{m-}^{k \theta}(\rho, \varphi, z) = \\
		&= \sqrt{\frac{k |\sin \theta|}{2\pi}} \iu^m \exp(\iu(k \cos \theta z + m\varphi)) \cdot \\
		&\cdot\left(\frac{\iu}{\sqrt{2}}\left[\left(1 + \cos \theta \right)J_{m+1}(k |\sin \theta| \, \rho) \exp(\iu\varphi) \mathbf{e}_L + \right. \right.\\
		&+ \left.
    \left(1 - \cos \theta \right) J_{m-1}(k |\sin \theta| \, \rho) \exp(-\iu\varphi) \mathbf{e}_R
    \right]
     \\ 
		&-\left.\frac{k |\sin \theta|}{k} J_m(k |\sin \theta| \, \rho) \mathbf{e}_z
    \right) \\ 
		&\mathcal{B}_{m+}^{k \theta}(\rho, \varphi, z) = \\ 
		&= \sqrt{\frac{k |\sin \theta|}{2\pi}} \iu^m \exp(\iu(k \cos \theta z + m\varphi)) \cdot \\
		&\cdot
    \left(
    \frac{\iu}{\sqrt{2}}
    \left[
    \left(
    1 - \cos \theta 
    \right) 
    J_{m+1}(k |\sin \theta| \, \rho) \exp(\iu\varphi) \mathbf{e}_L 
    + \right. \right.\\
		&+ \left.
    \left(1 + \cos \theta \right) J_{m-1}(k |\sin \theta| \, \rho) \exp(-\iu\varphi) \mathbf{e}_R
    \right]
     \\ 
		&+\left.\frac{k |\sin \theta|}{k} J_m(k |\sin \theta| \, \rho) \mathbf{e}_z
    \right).
\end{split}
\end{equation}
where $J_n(\cdot)$ are Bessel functions, and the polarization vectors $[\mathbf{e}_L,\mathbf{e}_R,\mathbf{e}_z]$ are defined in Tab.~\ref{tab:BB}.

The approximated forms of collimated Bessel beams can be obtained by taking the limit of small $|\theta|$: $\sin\theta\rightarrow\theta,\ \cos\theta\rightarrow1$, and then keeping the polarization with the largest coefficient multiplying the phase factors and Bessel functions.
For positive $k_z/k$ ($0\le \theta \le \pi/2$) we obtain:
\begin{equation}
	\label{eq:BBs_coll}
	\begin{split}
		&\mathcal{B}_{m-}^{k \theta\rightarrow 0}(\rho, \varphi, z) \rightarrow \\
		& \sqrt{\frac{k |\theta|}{\pi}} \exp(\iu k z ) 
		\iu^{m+1} J_{m+1}(k |\theta| \, \rho) \exp(\iu\varphi(m+1)) \mathbf{e}_L
     \\ 
		&\mathcal{B}_{m+}^{k \theta\rightarrow 0}(\rho, \varphi, z) \rightarrow\\ 
		& \sqrt{\frac{k |\theta|}{\pi}}  \exp(\iu k z ) \iu^{m+1} J_{m-1}(k |\theta| \, \rho) \exp(\iu\varphi(m-1)) \mathbf{e}_R
\end{split}
\end{equation}

\section{$\chi^{(3)}$ of isotropic nonlinear medium\label{app:chi3}}
\label{sec:tens}
In Cartesian coordinates, the $\chi^{(3)}_{ijkl}$ tensor of isotropic medium has the following form~\cite{Boyd2003}:
\begin{align}
    & \nonumber \chi^{(3)}_{xxxx}=\chi^{(3)}_{yyyy}=\chi^{(3)}_{zzzz}, \\
    &  \nonumber \chi^{(3)}_{xxyy}=\chi^{(3)}_{yyzz}=\chi^{(3)}_{zzxx}=\chi^{(3)}_{zzyy}=\chi^{(3)}_{yyzz}=\chi^{(3)}_{xxzz},\\ 
    & \nonumber \chi^{(3)}_{xzxz}=\chi^{(3)}_{xyxy}=\chi^{(3)}_{yzyz}=\chi^{(3)}_{yxyx}=\chi^{(3)}_{zyzy}=\chi^{(3)}_{zxzx},\\
    & \nonumber \chi^{(3)}_{xyyx}=\chi^{(3)}_{xzzx}=\chi^{(3)}_{yxxy}=\chi^{(3)}_{yzzy}=\chi^{(3)}_{zyyz}=\chi^{(3)}_{zxxz}.,\\
    & \chi^{(3)}_{xxxx} = \chi^{(3)}_{xxyy} + \chi^{(3)}_{xyxy} + \chi^{(3)}_{xyyx}.
\end{align} 
Let us provide the components with values $ 12, \ 4,\ 4, \ 4$ for $\chi^{(3)}_{xxxx}, \ \chi^{(3)}_{xxyy}, \ \chi^{(3)}_{xyxy}$ and $\chi^{(3)}_{xyyx}$, respectively. Then, in cylindrical coordinates we obtain~\cite{Nikitina2024-AchiralNanostructure}:
\begin{align}
    &\nonumber \chi^{(3)}_{zzzz} = 
    \chi^{(3)}_{\rho \rho \rho \rho} = \chi^{(3)}_{\varphi \varphi \varphi \varphi} = 12, \\
     &\chi^{(3)}_{\rho \rho \varphi \varphi} = \chi^{(3)}_{\rho \varphi  \rho \varphi} = \chi^{(3)}_{\rho \varphi \varphi \rho } =  \chi^{(3)}_{\varphi \rho \rho  \varphi} = \chi^{(3)}_{\varphi \varphi \rho \rho } = \nonumber
     \chi^{(3)}_{\varphi \rho \varphi \rho} =  \\ 
     = &\chi^{(3)}_{\varphi z \varphi z} = \chi^{(3)}_{\varphi z z \varphi } = \chi^{(3)}_{z \varphi z \varphi } = \chi^{(3)}_{z \varphi  \varphi z} = \chi^{(3)}_{\varphi \varphi z z} = \chi^{(3)}_{z z \varphi  \varphi } = \nonumber  \\
     = &\chi^{(3)}_{z \rho z \rho} = \chi^{(3)}_{ \rho z z \rho} = \chi^{(3)}_{\rho z \rho z} = \chi^{(3)}_{z z \rho  \rho} = \chi^{(3)}_{z \rho  \rho z} = \chi^{(3)}_{ \rho  \rho z z } = \nonumber \\ = &4.
\end{align}
One can see that all the components do not depend on $\varphi$, from which it follows that $m^\text{tens}=0$ (also by comparison with tensors for the lattices of another symmetry, given, e.g. in~\cite{Nikitina2024-AchiralNanostructure}).

\section{Theory of the TAM projection tripling in amorphous silicon}
\label{app:theorySi}
\subsection{Angular dependence}
\label{app:theorySiang}
Let us provide the theoretical description of the observed patterns using an alternative approach. 
In this section, we are only interested in azimuthal dependence of all functions, more precisely, in their symmetry behavior under rotations around the propagation ($z$-) axis.


In cylindrical coordinates $\left( \rho, \varphi, z \right)$, a circularly polarized plane wave  can be written as \cite{Frizyuk2021-NonlinearCircularDi}
\begin{equation}\label{eqn:circ_pol_plane_wave}
    \vb E^{\text{cp}} \propto (\vb e_\rho \pm \iu \vb e_\varphi) e^{\pm \iu \varphi},
\end{equation}
where $\pm$ corresponds to the right and left handedness, respectively.
The expression in Eq.~\eqref{eqn:circ_pol_plane_wave} has been chosen since the unit basis vectors of the cylindrical coordinate system do not change under rotation, and the exponential multiplier stands for the whole behavior.
Indeed, under rotation by an angle $\beta$ we just have $\varphi \to \varphi - \beta$.
{If the pump beam has arbitrary radial dependence and right handedness}, the electric field takes the form $\mathbf{E}^{\omega}(\rho, \varphi) = \vb{\tilde E}^{\omega}(\rho)e^{\iu \varphi}$,
thus from Eq.~\eqref{polar_3} for the nonlinear polarization, since $m^\text{tens}=0$, we have:
\begin{equation}
    \vb P^{3\omega}_0(\rho, \varphi) \propto \vb {\tilde P}^{3\omega}_0(\rho) e^{3 \iu \varphi},
\end{equation}
where $\vb P^{3\omega}_0(\rho)$ does not depend on $\varphi$, i.e. it is invariant under rotations. 
The intensity distribution, produced by this source, will possess the same symmetry behavior.
However, if we the {helicity} of the pump is not {pure}, i.e. there is a contribution with the opposite handedness (opposite TAM projection), $\varepsilon e^{- \iu \varphi}$, we will obtain additional terms
\begin{align}
    \vb P^{3\omega}_1(\rho, \varphi) &\propto \varepsilon \vb {\tilde P}^{3\omega}_1(\rho) e^{1 \iu \varphi}\\
    \vb P^{3\omega}_2(\rho, \varphi) &\propto \varepsilon^2 \vb {\tilde P}^{3\omega}_2(\rho) e^{-1 \iu \varphi}\\
    \vb P^{3\omega}_3(\rho, \varphi) &\propto \varepsilon^3 \vb {\tilde P}^{3\omega}_3(\rho) e^{-3 \iu \varphi}.
\end{align}
Now we consider the angular dependence of the total TH intensity.
Taking into account only the fields produced by the first two terms, i.e. $\vb P^{3\omega}_0(\rho, \varphi)$ and $\vb P^{3\omega}_1(\rho, \varphi)$, and neglecting the terms of higher order, we obtain 
\begin{align} \label{eqn:I_3w_01}
    I^{3\omega} \propto |(\vb a(\rho) e^{3 \iu \varphi} +  \vb b(\rho) \varepsilon e^{1 \iu \varphi})|^2 \propto c(\rho) + d(\rho) \cos (2\varphi).
\end{align}
where the vector functions $\vb a(\rho)$ and $\vb b(\rho)$ characterize the radial dependence of polarization.  Equation~\eqref{eqn:I_3w_01} indicates that, in general, the TH signal generated by a nonpure pump represents a two-lobe pattern (see Fig.~\ref{spiral_exp&sim}). 
Radial dependence, described by the functions $\vb a, \vb b, c, d$ will be discussed in more detail in Section~\ref{sec:radial}. 
\subsection{Radial dependence}
\label{sec:radial}

Now we focus on the radial dependence of the formed patterns. On the sample plane, one may decompose the incident wave into the terms with right, left, and longitudinal polarization.
To this purpose, the following terms can be introduced 
\begin{align}
\label{pol_vec1}
    \frac{1}{\sqrt{2}}(\xhat + \iu\yhat) = \frac{1}{\sqrt{2}}(\vb e_\rho + \iu \vb e_\varphi)e^{\iu\varphi} &=  \vb e_R \\
    \label{pol_vec2}
    \frac{1}{\sqrt{2}}(\xhat - \iu\yhat) = \frac{1}{\sqrt{2}}(\vb e_\rho - \iu \vb e_\varphi)e^{-\iu\varphi} &=  \vb e_L \\
    \vb e_z &=  \vb e_z
    \label{pol_vec3}
\end{align}
Because of rotational symmetry of the focusing system, the focused electric field should contain only terms possessing the same $m$, as the input electric field.
Thus, from Eqs.~\eqref{pol_vec1}-\eqref{pol_vec3} and assuming that they can have arbitrary radial dependence, we can deduce the form for a focused field with $m = 1$ in the most general case: 
\begin{align}
    \vb{E}^\text{f}(r,\varphi,z) = f_0\vb e_R + f_2e^{2\iu\varphi}\vb e_L + f_1e^{\iu\varphi}\vb e_z.
    \label{pump_eq}
\end{align}
The exponential terms in Eq.~\eqref{pump_eq} appear to save the rotational symmetry. Therefore, it follows that $f_0(\rho, z)$ will be non-zero in center, while $f_1(\rho, z)$ and $f_2(\rho, z)$ should vanish, because the angular part has phase singularity. 

\subsubsection{On particular form of radial dependence in our case}
\label{sec:foc_pw_an} 
 For a RCP plane wave at the input, the focused field is given by~\cite{richards1959electromagnetic, Kant1993-AnAnalyticalSolutio}:
\begin{align}
    \mathbf{E}^\text{f}(\rho,\varphi, z) \propto  \frac{1}{\sqrt{2}}(I_0\mathbf{e}_R + I_2e^{ 2 \iu \varphi}\mathbf{e}_L) - \iu I_1e^{\iu \varphi}\mathbf{e}_z,
    \label{eq:foc_orig}
\end{align}
where $(\rho, \varphi, z)$ are the coordinates of the observation point in a cylindrical coordinate system (see  Fig.~\ref{pump_sch}a) and the functions $I_i(\rho, z)$, with $i=0,1,2$,  are expressed in integral form and explicitly are given as follows:
\begin{align}
\label{eq:pw1}
    I_0(\rho, z)  = &\int_{0}^{\theta_{\text{max}}} g(\theta)\sin\theta(1+\cos\theta)\cdot\\
    & \nonumber \cdot J_0(k\rho\sin\theta)\exp(\iu kz\cos\theta) \dd\theta \\
    I_1(\rho, z)  = &\int_{0}^{\theta_{\text{max}}} g(\theta)\sin^2(\theta)\cdot 
    \\
    & \nonumber \cdot J_1(k\rho\sin\theta)\exp(\iu kz\cos\theta) \dd\theta\\
    \label{eq:pw3}
    I_2(\rho, z)  = &\int_{0}^{\theta_{\text{max}}} g(\theta)\sin\theta(1-\cos\theta)\cdot     \\&\cdot J_2(k\rho\sin\theta)\exp(\iu kz\cos\theta) \dd\theta \nonumber
\end{align}
with $g(\theta)$ being the apodization factor, which for an aplanatic lens is equal to $\sqrt{\cos(\theta)}$, and $\theta_{\text{max}}$ = asin(NA), where NA is the objective lens numerical aperture, $k = \omega/c$.

\subsection{Polarization and electric field distribution at TH generated by a focused plane wave}
From Eq.~\eqref{polar_3} and taking into account the intrinsic permutation symmetry~\cite{Boyd2003} for components of the polarization field induced in a thin isotropic layer at 3$\omega$, we have:
\begin{align}
\nonumber
    P_x^{3\omega} \propto E_{x}E_{x}E_{x} + E_{x}E_{y}E_{y} + E_{x}E_{z}E_{z} \\
    \nonumber
    P_y^{3\omega} \propto E_{y}E_{y}E_{y} + E_{y}E_{x}E_{x} + E_{y}E_{z}E_{z}\\
    P_z^{3\omega} \propto E_{z}E_{z}E_{z} + E_{z}E_{y}E_{y} + E_{z}E_{x}E_{x}
\end{align}

Therefore, the components of the polarization induced by the focused RCP plane wave in Eq.~\eqref{pump_eq} are given by:
\begin{align}
\nonumber
    P_x^{3\omega} &\propto (4f_2^2f_0 + f_2f_1^2)e^{4\iu\varphi} + (4f_2f_0^2 + f_0f_1^2)e^{2\iu\varphi} \\
    \nonumber
    P_y^{3\omega} &\propto -\iu(4f_2^2f_0 + f_2f_1^2)e^{4\iu\varphi} + \iu(4f_2f_0^2 + f_0f_1^2)e^{2\iu\varphi}\\
    P_z^{3\omega} &\propto (f_1^3 + 4f_2f_0f_1)e^{3\iu\varphi} 
\end{align}
Taking into account Eqs.~\eqref{pol_vec1}-\eqref{pol_vec3}, the polarization at $3\omega$ takes the form:
\begin{align}
\nonumber
    \mathbf{P}^{3\omega}(\rho, \varphi, z) &\propto (4f_2f_0^2 + f_0f_1^2)e^{2\iu\varphi}\mathbf{e}_R + \\ \nonumber  &+ (4f_2^2f_0 + f_2f_1^2)e^{4\iu\varphi} \mathbf{e}_L +\\
    &+(4f_0f_2f_1 + f_1^3)e^{3\iu\varphi} \mathbf{e}_z, 
    \label{polar_3om}
\end{align}
which can be rewritten as:
\begin{align}
\mathbf{P}^{3\omega}(\rho, \varphi, z) = \mathbf{\tilde{P}}^{3\omega}(\rho, z)e^{3\iu\varphi}.
\label{polar_tripl}
\end{align}
At this stage it is worthy to note that the presence of exponential multiplier in Eq.~\eqref{polar_tripl} is consistent with prediction of the general theory, confirming the tripling of total orbital momentum projection.
On the other hand, the polarization induced by an RCP plane wave with LCP contribution proportional to $\varepsilon $, is given as follows:
\begin{align}
\nonumber
    \mathbf{P}^{3\omega}(\rho, \varphi, z) &\propto g_0e^{2\iu\varphi}\mathbf{e}_R + g_2e^{4\iu\varphi} \mathbf{e}_L +g_1e^{3\iu\varphi} \mathbf{e}_z + \\
    \nonumber & +\varepsilon(h_0\mathbf{e}_R + h_2 e^{2\iu\varphi}\mathbf{e}_L + h_1 e^{\iu\varphi} \mathbf{e}_z) +\\
    \nonumber & +\varepsilon^2(h_2e^{-2\iu\varphi}\mathbf{e}_R + h_0 \mathbf{e}_L + h_1 e^{-\iu\varphi} \mathbf{e}_z) +\\
     & +\varepsilon^3(g_2e^{-4\iu\varphi}\mathbf{e}_R + g_0 e^{-2\iu\varphi}\mathbf{e}_L + g_1 e^{-3\iu\varphi} \mathbf{e}_z) 
    \label{polar_3om_eps}
\end{align}
where we introduced $g_0 = 4f_2f_0^2 + f_0f_1^2$, $g_1 = 4f_2f_0f_1 + f_1^3$, $g_2 = 4f_2^2f_0 + f_2f_1^2$, $h_0 = 4f_0^3 + 8f_2^2f_0 + 2f_0f_1^2 + f_2f_1^2$, $h_1 = 4f_2f_0f_1 + 3f_1^3 + 4f_2^2f_1+4f_0^2f_1$ and $h_2 = 4f_2^3+8f_2f_0^2+2f_2f_1^2+f_0f_1^2$.

One can note that polarization induced by a non-pure pump field can not be represented in the form of equality \eqref{polar_tripl}, since it contains terms with $m^{3\omega} \in \{3, 1, -1, 3\}$, as also shown in Eq.~\eqref{eq:mm}. 

\begin{figure}[h!]
    \centering
    \includegraphics[width=\columnwidth]{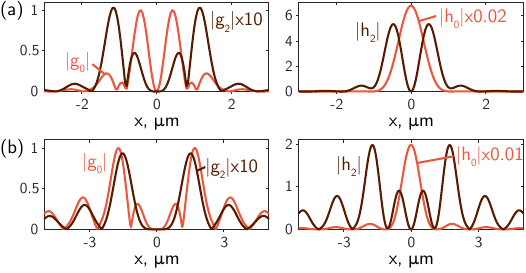}
    \caption{Calculated normalized components of polarization $\mathbf{P}^{3\omega}(z_0=0)$ with pump being (a) a focused plane wave (NA = 0.85) , and (b) a Bessel beam (1,1) with $\theta = \asin(0.5)$. } 
    \label{polar_res22}
\end{figure}
{
Figure~\ref{polar_res22} shows the radial behavior of functions $g_0$, $g_2$, $h_0$, $h_2$ for the case of a focused plane wave excitation (Eq.~\eqref{eq:foc_orig})
and a pure Bessel beam
$(1, 1)$, with $f_0 \propto (1+\cos(\theta))J_0, \ f_1\propto\sin(\theta)J_1, \ f_2\propto (1-\cos(\theta))J_2$. 
The left panel of Fig.~\ref{polar_res22} shows that the $\rhat$ component, $|g_0|$ (orange curves), is dominant for the terms of $\vb P^{3\omega}$, which do not depend on $\varepsilon$. It is expected from other $\theta$-dependent factors multiplying each polarization in Eq.~\eqref{eq:BBs}, namely $(1+\cos\theta)$ for $\rhat$ and  $(1-\cos\theta)$ for $\lhat$, for the case of Bessel beams, and more complicated Eq.~\eqref{eq:pw1}-Eg.~\eqref{eq:pw3} for the focused plane wave. 
Therefore, dominating contribution of $\varepsilon$-independent terms of TH response is $(3, 1)$, and for $\varepsilon^3$-proportional terms it is $(-3, -1)$.
One can see that for a Bessel beam only $h_0$ (orange curves) would have maximum in the center, which coincides with the results for a plane wave, depicted in~Fig.~\ref{polar_res22}a.
The right panel in Fig.~\ref{polar_res22} corresponds to the terms, proportional to $\varepsilon$, where the contribution of $(1, 1)$ is dominant, and to $\varepsilon^2$, with dominance of $(-1,-1)$.}

\section{Expression for $\varepsilon(\gamma)$}

\label{app:vareps_circ}
Electric field $\vb E^\prime$, transmitted through a wave plate with a given retardance angle $\tau$ and the angle $\gamma$ between the $y$-axis and the fast axis can be written in terms of Jones vectors as follows~\cite{Shurcliff+1962, collett1992polarized}:
\begin{align}
\label{eq:e_retard}
    \begin{pmatrix} E_x^\prime \\ E_y^\prime \end{pmatrix} = R(-\gamma)M(\tau)R(\gamma)
    \begin{pmatrix} E_x \\ E_y \end{pmatrix}
\end{align}
where
\begin{align}
 M(\tau) = \begin{bmatrix} 
 \exp(\iu \tau) & 0 \\ 0 & 1
 \end{bmatrix}
\end{align}
and
\begin{align}
 R(\gamma) = \begin{bmatrix} 
 \cos\gamma & -\sin\gamma \\ \sin\gamma & \cos\gamma
 \end{bmatrix}
\end{align}
From Eq. (\ref{eq:e_retard}) for $y$-polarized input $\vb E = (0, 1)^\mathsf{T}$ we have:
\begin{align}
\label{eq:E_ret_xy}
   \nonumber \vb{E}^\prime = (1-\exp(\iu\tau))\sin\gamma\cos\gamma\xhat + \\
    +(\sin^2\gamma\exp(\iu \tau) + \cos^2\gamma)\yhat
\end{align}
Taking into account (\ref{pol_vec1}) and (\ref{pol_vec2}), we can rewrite $\vb{E}^\prime$ in the form $\vb e_R + \varepsilon \vb e_L$:
\begin{align}
    &\nonumber \vb{E}^\prime \propto
    \\
    &\nonumber[ -\iu(\sin^2\gamma\exp(\iu\tau) + \cos\gamma^2 ) + (1-\exp(\iu\tau))\sin\gamma\cos\gamma]\vb e_R+\\
    &\nonumber+[ \iu(\sin^2\gamma\exp(\iu\tau) + \cos^2\gamma) + (1-\exp(\iu\tau))\sin\gamma\cos\gamma]\vb e_L
\end{align}
Finally, dividing by the coefficient in front of $\rhat$ and simplifying for $\varepsilon$ we get:
\begin{equation}
    \varepsilon(\gamma) = -\frac{\iu\sin2\gamma \cos\tau + \cos 2\gamma}{\sin2\gamma \sin\tau + 1}e^{-2\iu \gamma}
\end{equation}
From this expression we can find the ellipticiy angle $\alpha$ and the ellipse inclination angle $\beta$ in terms of $\gamma$ and $\tau$:
\begin{align}
\label{eq:alph_bs}
    \sin2\alpha = \frac{S_3}{S_0} = \frac{1-|\varepsilon|^2}{1+|\varepsilon|^2} =\sin2\gamma \sin \tau\\
    \tan2\beta = \frac{S_2}{S_1} =\frac{\text{Im}\varepsilon}{\text{Re}\varepsilon} = \frac{-\sin2\gamma\cos2\gamma(1-\cos\tau)}{\cos\tau\sin^22\gamma + \cos^22\gamma}
    \label{eq:beta_bs}
\end{align}
Equations~\eqref{eq:alph_bs}, \eqref{eq:beta_bs} can be obtained considering the Mueller matrix of a Babinet-Soleil compensator with a $y$-polarized input~\cite{collett1992polarized}.

\onecolumn
\section{Schematic of the optical setup} 
\label{app:setup}
\begin{figure*}[h!]
    \centering
    \includegraphics[width=\textwidth]{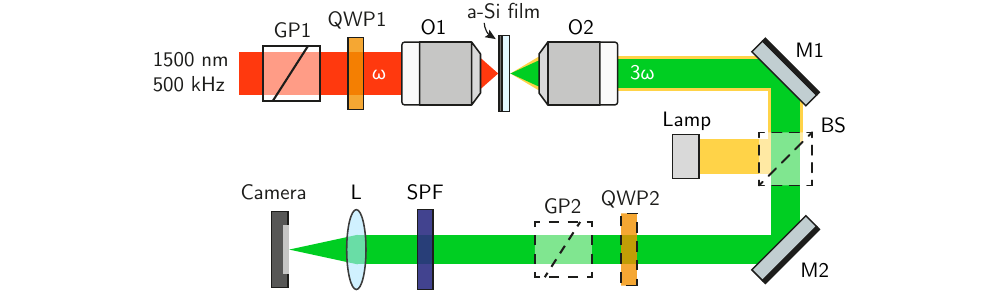}
    \caption{Schematic of the optical setup; here GP --- Glan polarizer, M --- mirror, L --- lens, SPF --- short pass filter, O --- objective, BS --- beam splitter.  } 
    \label{setup}
\end{figure*}

Figure \ref{setup} shows a schematic of the optical setup. The elements GP2 and QWP2 were used when conducting polarization-resolved measurements. The removable beam splitter (BS) was used only to illuminate the sample when focusing on the surface of the a-Si film. 

\section{Numerical simulation of TH generation with and without the substrate}
\label{app:sim_subs}
\begin{figure*}[h!]
    \centering
    \includegraphics[width=\textwidth]{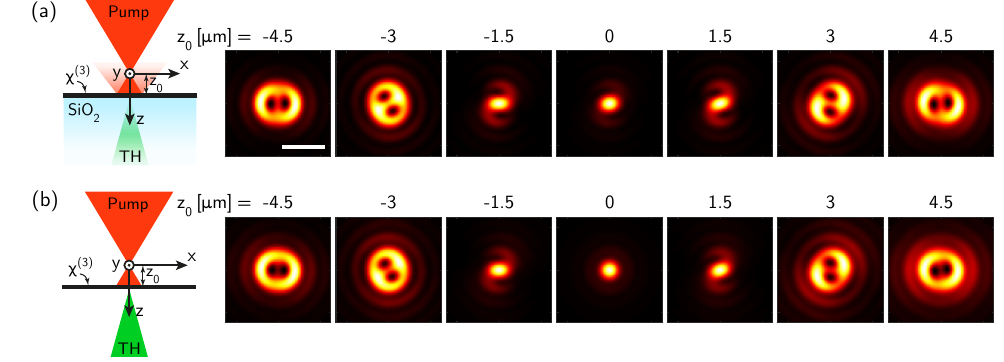}
    \caption{a) TH intensity patterns calculated numerically, taking into account the presence of the air/substrate interface, b) patterns calculated for the nonlinear layer placed in the air; $z_0$ is the defocusing parameter. Scale bar, 2 $\mu$m. } 
    \label{subs_compar}
\end{figure*}

Figure \ref{subs_compar} shows a numerical simulation of TH generation by a focused laser beam when either taking into account the presence of the air/silica substrate interface or neglecting it. 
Comparing the upper and bottom panels, one can see that the presence of the interface does not significantly alter the TH patterns. Therefore, for simplicity, in the main text we discuss the case without the air/substrate interface, shown in Fig. \ref{subs_compar}b.

\section{Background TH generation by the optics of the setup} 

\begin{figure}[h]
    \centering
    \includegraphics[width=\columnwidth]{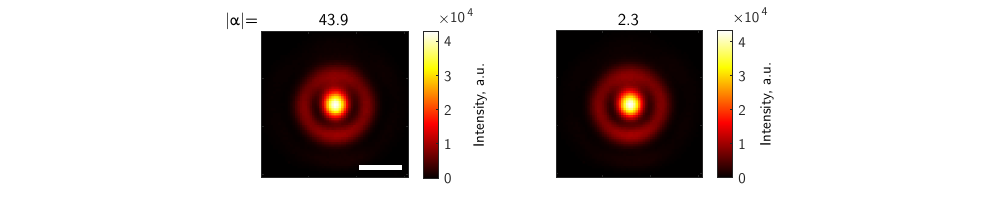}
    \caption{TH signal collected when focusing on substrate/air interface with ellipticity angle of the pump equal to 43.9$^\circ$ (close to circular) and 2.3$^\circ$ (close to linear). TH signal shape and intensity does not depend on the pump polarization. Scale bar, 2 $\mu$m. } 
    \label{bg_thg}
\end{figure}

Figure \ref{bg_thg} shows experimentally measured TH signals, when focusing on the silica substrate/air interface. One can see that in this case the signal intensity and its shape do not depend on the polarization of the pump. This behavior can be explained by the generation of parasitic TH in the optical elements of the setup. When focusing on the film, this signal is strongly attenuated, due to the high absorption of a-Si at the TH frequency ($>$40 dB/$\mu$m at 500 nm \cite{Hishikawa_1991}),

\section{Polarization resolved measurements of TH signal } 

\begin{figure*}[h!]
    \centering
    \includegraphics[width=\textwidth]{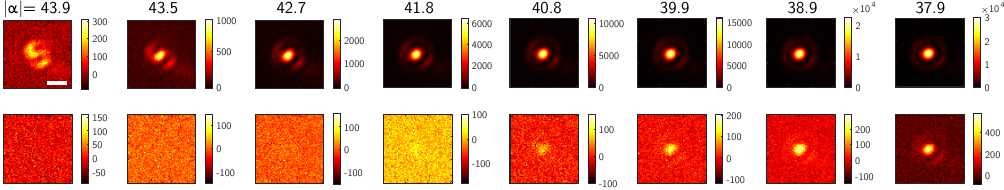}
    \caption{Polarization resolved measurements of TH signal, obtained when probing RCP (top panel) and LCP (bottom panel) light. Scale bar, 2 $\mu$m.} 
    \label{polar_res_all}
\end{figure*}

Figure \ref{polar_res_all} shows additional intensity patterns of the TH signal obtained in the polarization resolved experiment. In Fig.~(\ref{polar_res}) of the main text are shown patterns obtained with pump ellipticity angles $\alpha$ of $43.9^\circ$ and $37.9^\circ$. 

\end{document}